\documentclass[aps,reprint]{revtex4-1}

\usepackage{amssymb,amsmath} 
\usepackage{graphicx}

\usepackage{appendix}
\usepackage[small,bf]{caption}
\usepackage{rotating}

\usepackage{color}
\usepackage[usenames,dvipsnames,svgnames,table]{xcolor}
\begin{document}

\author{Rhiannon Pinney}
\affiliation{HH Wills Physics Laboratory, Tyndall Avenue, Bristol, BS8 1TL, UK}
\affiliation{Bristol Centre for Complexity Science, University of Bristol, Bristol, BS8 1TS, UK}

\author{Tanniemola B.  Liverpool}
\affiliation{School of Mathematics, University of Bristol, Bristol, BS8 1TW, UK}

\author{C. Patrick Royall}
\affiliation{HH Wills Physics Laboratory, Tyndall Avenue, Bristol, BS8 1TL, UK}
\affiliation{School of Chemistry, University of Bristol, Cantock Close, Bristol, BS8 1TS, UK}
\affiliation{Centre for Nanoscience and Quantum Information, Tyndall Avenue, Bristol, BS8 1FD, UK}

\email{paddy.royall@bristol.ac.uk}

\title{Recasting a model atomistic glassformer as a system of icosahedra}

\begin{abstract}
We consider a binary Lennard-Jones glassformer whose super-Arrhenius dynamics are correlated with the formation of icosahedral structures. Upon cooling these icosahedra organize into \textcolor{black}{\emph{mesoclusters}}. We recast this glassformer as an effective system of icosahedra which we describe with a population dynamics model. This model we parameterize with data from the temperature regime accessible to molecular dynamics simulations. We then use the model to determine the population of icosahedra in mesoclusters at arbitrary temperature. Using simulation data to incorporate dynamics into the model we predict relaxation behavior at temperatures inaccessible to conventional approaches. Our model predicts super-Arrhenius dynamics whose relaxation time remains finite for non-zero temperature.
\end{abstract}


\maketitle

%\textcolor{black}{\section{Introduction}}
\section{Introduction}
\label{sectionIntroduction}

Among the challenges of the glass transition is to understand how solidity emerges with little apparent change in structure. Indeed, whether the glass transition has a thermodynamic (implying structural) or  dynamical origin remains unclear  \cite{cavagna2009,berthier2011}. It has been proposed that upon cooling, icosahedral arrangements of atoms might form in supercooled liquids \cite{frank1952} and that dynamical arrest may be related to a (geometrically frustrated) transition to a phase of such icosahedra \cite{tarjus2005}. It is now possible to identify geometric motifs such as icosahedra and related \emph{locally favored structures} (LFS) using computer simulation \cite{steinhardt1983,jonsson1988,dzugutov2002,coslovich2007,eckmann2008,lerner2009,sausset2010,tanaka2010,malins2013jcp,hocky2014,royall2014} and particle-resolved studies in colloid experiments \cite{ivlev,royall2008,mazoyer2011,leocmach2012}. Further evidence of increasing numbers of LFS upon cooling is also found in metallic glassformers \cite{royall2015physrep,cheng2011}.

The discovery of dynamic heterogeneity, \emph{i.e.} locally fast and slow regions \cite{hurley1995,ediger2000} has spurred attempts to correlate LFS such as icosahedra with the dynamically slow regions. This has met with some success \cite{dzugutov2002,coslovich2007,tanaka2010,sausset2010,malins2013jcp,jack2014,hocky2014} however such correlation does not by itself demonstrate a mechanism for arrest \cite{charbonneau2013pre,jack2014} and in any case is dependent on the model system under consideration \cite{hocky2014}. A key limitation here is that direct detection of LFS and dynamic heterogeneity is only possible in the first 4-5 decades of dynamic slowing accessible to particle-resolved experiments on colloids \cite{ivlev} and computer simulation. This compares to 14 decades of slowing corresponding to the glass transition in molecular systems and divergence of relaxation time at a \textcolor{black}{putative} thermodynamic transition. Clearly, any discussion of the nature of the glass transition itself requires extensive extrapolation of data. Significantly, the limit of this accessible 4-5 decades corresponds roughly to the Mode-Coupling temperature $T_\mathrm{MCT}$ which in $d=3$ leads to a crossover to a regime where relaxation is believed to occur in a qualitatively different fashion through cooperative behavior \cite{berthier2011,lubchenko2007,cavagna2012}.

Recently it has become possible to access certain properties of this deeply supercooled regime. One approach is to vapor deposit onto a substrate cooled below the temperature at which the system can usually be equilibrated \cite{swallen2007,singh2013}.  Alternatively, by immobilizing or ``pinning'' a subset of particles a transition to an ``ideal glass'' can be induced that is accessible to simulation \cite{cammarota2012,kob2013} and experiments with colloids \cite{gokhale2014}. Other methods include the observation of a transition in distributions of overlaps in configuration space \cite{berthier2013overlap,berthier2015} and so-called large deviations where trajectory sampling of moderately supercooled liquids indicates a dynamical transition to a state rich in LFS with very slow dynamics \cite{hedges2009,speck2012}. \textcolor{black}{It is also possible to} decompose the system into a range of geometric motifs.  \textcolor{black}{Such an approach} indicated that there is no thermodynamic transition \cite{eckmann2008,lerner2009}. However obtaining dynamical information in the deeply supercooled regime \textcolor{black}{beyond that accessible to simulation} remains a challenge.

\begin{figure}%[!htb]
\begin{center}
\includegraphics[width=70mm]{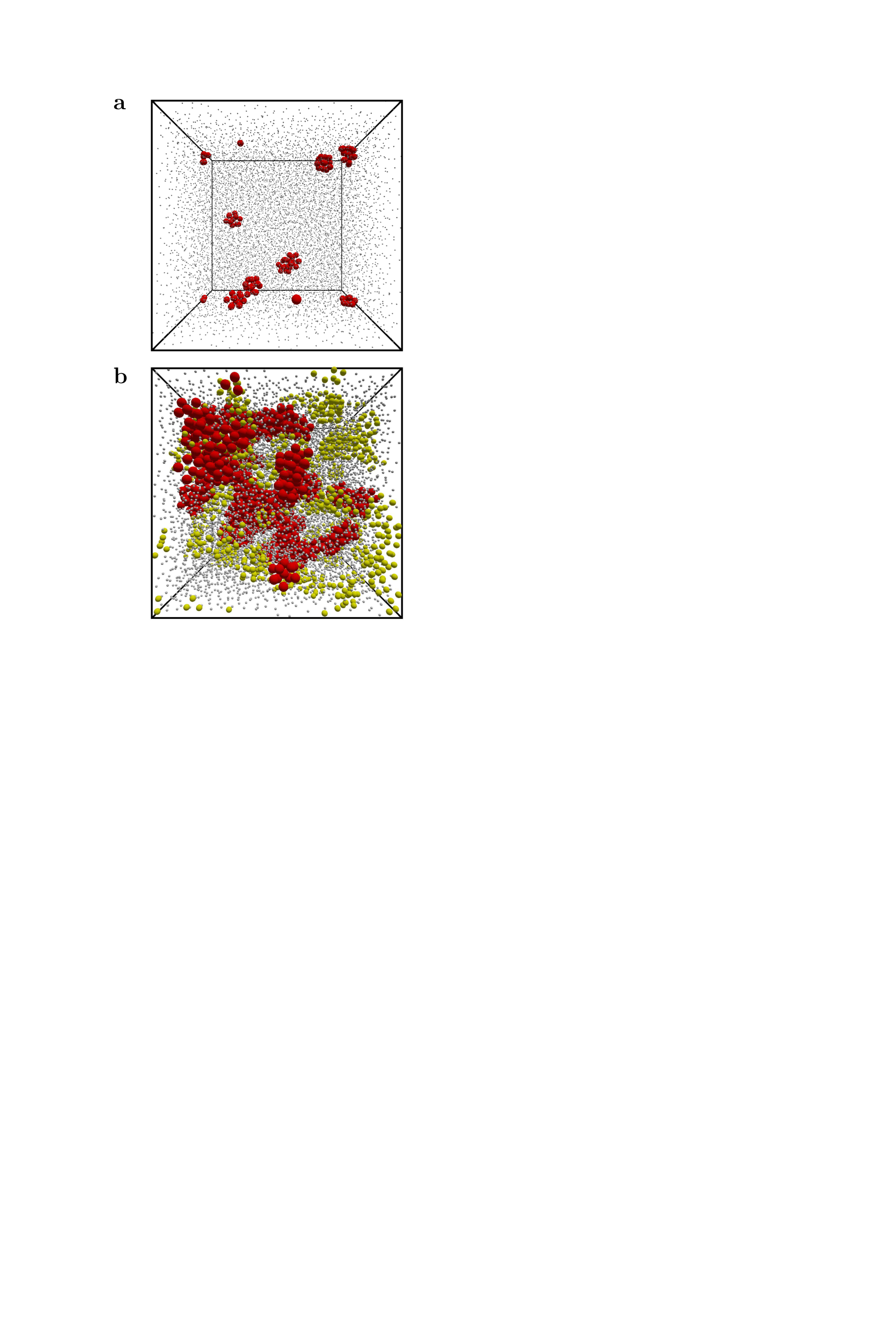}
\caption{(Color online) Snapshots showing the simulated system. Large, colored particles indicate those identified as being part of icosahedra. Small, gray particles are all other particles. Sizes are not to scale. (a) At high temperatures ($T = 1.0$), we see very few, small mesoclusters and (b) at low temperatures ($T = 0.58$) a network of very large mesoclusters forms (red particles). These mesoclusters percolate at around $T = 0.6$.}
\label{figSnapshot}
\end{center}
\end{figure}

\begin{figure}%[!htb]
\begin{center}
\includegraphics[width=85mm]{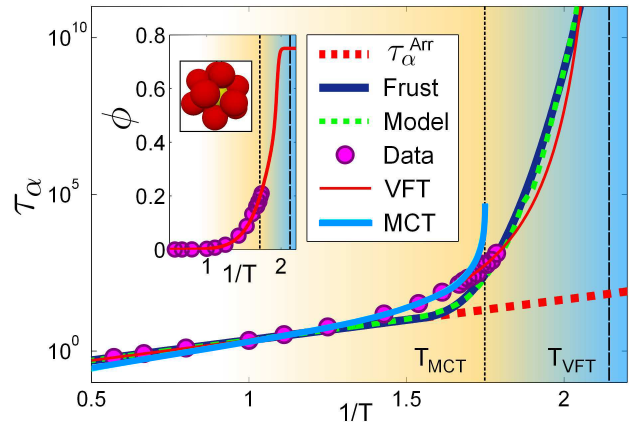} 
\caption{(Color online) ``Angell'' plot of relaxation time as a function of inverse temperature. Circles; simulation data. Dashed red line; Arrhenius. Dashed green line; population dynamics model prediction. Solid navy line; as described by Geometric Frustration \cite{tarjus2005}. Thinner, solid pale blue line; MCT fit to the data across the region $0.58 \leq T \leq 1$. Thin, solid red line; VFT fit to data $T \leq 1$. Inset: proportion of particles identified as being part of an icosahedra as a function of temperature. Circles; simulation data. Solid red line; fitted model. Simulations are limited to $T \gtrsim 0.56$ (blue shaded region indicates inaccessibility). Black dashed lines corresponds to $T_\mathrm{VFT}=0.456$ and $T_\mathrm{MCT}=0.57$.}
\label{figsAngellAndPhiAll}
\end{center}
\end{figure}

Here, we consider a binary Lennnard-Jones glassformer whose dynamics are strongly correlated with LFS, which are icosahedra \cite{coslovich2007,malins2013jcp,hocky2014}. In particular, the occurrence of super-Arrhenius dynamics coincides with the emergence of a population of icosahedra  (Figs. \ref{figSnapshot} and \ref{figsAngellAndPhiAll}) \cite{coslovich2007,malins2013jcp}. We build on this observation and use the population of icosahedra and its dynamics to predict the behavior in the regime beyond that accessible to computer simulation. To do this, we introduce a stochastic model for the population dynamics of icosahedra which we parameterize in the regime accessible to simulation. We then use the model to obtain the population of icosahedra for all temperatures. In the simulation accessible regime we show that the population and lifetime of the domains of icosahedra \textcolor{black}{which we term \emph{mesoclusters}} is \textcolor{black}{correlated with} the super-Arrhenius dynamics. Using the calculated population of icosahedra we predict the dynamical behavior of the system at all temperatures. \textcolor{black}{In particular we make the following assumptions: the dynamical behaviour of the system is democratically represented by particles in different sized mesoclusters of icosahedra and those not in icosahedra. The dynamics of particles not in icosahedra we assume to be Arrhenius. The dynamics of particles in mesoclusters we determine from parameterisation of mesocluster lifetimes using simulation.} The model predicts a super-Arrhenius dynamics comparable to the simulation data, however it indicates there is no divergence of the relaxation time at finite temperature.

Our model captures the super-Arrhenius behavior of the system with reasonable accuracy. The largest discrepancy between the predictions of the model and the simulation data occurs during the first few decades of super-Arrhenius dynamics around $0.7\gtrsim T\gtrsim0.6$. These first few decades of arrest are well-described by Mode-Coupling theory. MCT takes as its input two-point correlations. These are entirely neglected in our analysis which focuses on higher order correlations, and we attribute the discrepancy \textcolor{black}{predominantly} to our neglect of MCT.

This paper is divided as follows: we discuss the simulation details in section \ref{sectionSimulation}, and describe our model in section \ref{sectionModel}. Results are shown in section \ref{sectionResults}, and we conclude with a summary and discussion in section \ref{sectionSummary}.

%\textcolor{black}{\section{Simulation details}}
\section{Simulation details} 
\label{sectionSimulation}

\begin{figure}%[!htb]
\begin{center}
\includegraphics[width=80mm]{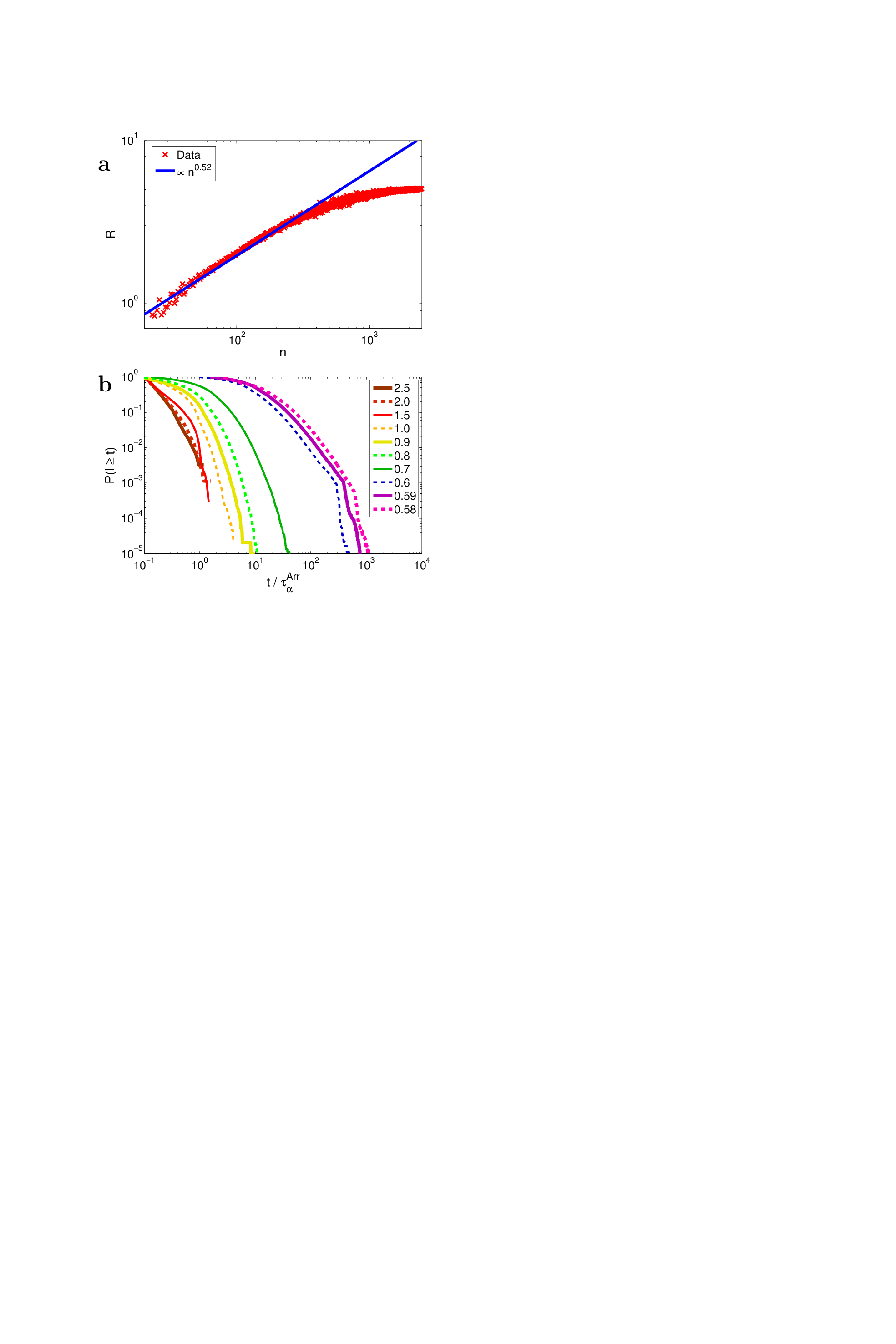} 
\caption{(Color online) (a) The radius of gyration for different size mesoclusters (here, size is determined by number of particles). Data (red crosses) fitted with $R^{2} \propto n^{1/d}$, where $d$ is the fractal dimension, and the fitted value $1/d = 0.52$; $d = 1.92$. (b) Distribution of icosahedra lifetimes for a range of temperatures, scaled by the Arrhenius timescale, $\tau_{\alpha}^\mathrm{Arr}$.}
\label{figFractalDimension}
\end{center}
\end{figure}

We simulate the Wahnstr\"{o}m equimolar additive binary Lennard-Jones model \cite{wahnstrom1991}. The size ratio is $5/6$ and the well depth between all species is identical. The mass of the large particles is twice that of the small. We use molecular dynamics  simulations of $N=1372, 10976, 87808$ particles. We equilibrate for at least $100 \tau_\alpha$ in the NVT ensemble before sampling in the NVE ensemble. Here $\tau_\alpha$ is the structural relaxation time which is determined from a stretched exponential fit to the intermediate scattering function. We identify icosahedra with the topological cluster classification (TCC) and consider those which last longer than $0.1\tau_\alpha$ (the distributions of which can be seen in Fig. \ref{figFractalDimension}(a)) to suppress the effects of thermal fluctuations. Our simulation and analysis protocol are detailed in \cite{malins2013jcp,malins2013tcc}.

\textcolor{black}{The Wanstr\"{o}m mixture has been shown to crystallise to a Frank-Kasper phase \cite{pedersen2010}. Indeed some simulations of $N=1372$ particles crystallised at temperatures $T \leq 0.6$; clearly evidenced by a substantial increase in the population of icosahedra at fixed temperature. These simulations were discarded. No crystallisation was observed in the larger systems.}

We plot the structural relaxation time $\tau_\alpha$ with an Arrhenius form for $T \gtrsim 1$ and a Vogel-Fulcher-Tamman (VFT) form for lower temperatures in Fig. \ref{figsAngellAndPhiAll}. The VFT form reads $\tau_\alpha=\tau_0 \exp[D/(T-T_0)]$ where the fragility parameter $D=0.799$ and the temperature at which our fit predicts a divergence of $\tau_\alpha$ at a temperature $T_0=0.456$. In Fig.  \ref{figsAngellAndPhiAll} we also indicate the Mode-Coupling temperature $T_\mathrm{MCT}=0.57$, fitted across the region $0.58 \leq T \leq 1$.

%\textcolor{black}{\section{Population dynamics model}} 
\section{Population dynamics model}
\label{sectionModel}

%\textcolor{black}{\subsection{}}
%\subsection{A global population of icosahedra}

In Fig. \ref{figSnapshot}, we see domains of icosahedra which we term \emph{mesoclusters}. We define the size of a mesocluster, $m$, by the number of icosahedra that comprise it; i.e. how many particles found at the center of an icosahedron are contained in the mesocluster; so here we only concern ourselves with $m \geq 1$ ($m = 0$ refers to particles that are not in an icosahedron).

We make the assumption that any change in mesocluster size is a change in $m$ of $\pm 1$. \textcolor{black}{This assumption means that even large mesoclusters (which in practise might break in two) can only decrease by incrementally. Thus mesocluster scission, or indeed coalescence, is not a feature of our model.} The rates at which the mesocluster size increases or decreases are given by $g$ and $r$ respectively. The master equation for the population dynamics model reads:

\begin{align}
\dot{p}_{1} &= g_0 p_{0} + r_2 p_{2} - [g_1+r_1]p_1 \nonumber \\
\dot{p}_{m} &= g_{m-1}p_{m-1} + r_{m+1}p_{m+1} - [g_m+r_m]p_{m} \nonumber \\
\dot{p}_{M} &= g_{M-1}p_{M-1} - r_Mp_{M}
\label{eqMaster}
\end{align}

\noindent where $p_{m}$ is the probability that a particle is in a mesocluster of size $m$. We set a limit at $m=M$ which is the largest possible mesocluster that can be formed given the relative population of particles in icosahedra, $\phi$. Considering geometric frustration we expect $\phi<1$ and set \textcolor{black}{a maximum value for the relative population $\phi_\mathrm{max}=0.75$.} This in turn constrains the largest number of icosahedra in a mesocluster, $M$. While the limit of the parameter $\phi$ is chosen, rather than determined, results obtained in the range $0.6 \leq \phi_\mathrm{max} \leq 0.9$ exhibit only slight quantitative differences and have no impact on our conclusions.

The steady state solution for the master equation (solution when $\dot{p}_{m}=0$ for all $m$) is:

\begin{equation}
p_{m} = \frac{g_{m-1}}{r_m}p_{m-1} = \frac{g_1 \cdots g_{m-1}}{r_2 \cdots r_m}p_{1}
\label{eqSSSolution}
\end{equation}

\noindent If $g_m, r_m$ are assumed constant across all $m$ (for a given $T$), $g_m = g$ and $r_m = r$, we can denote the ratio as a ``decay parameter'' $a = g/r$. This results in $p_m = a^{m-1}p_1$, so all $p_m$ can be determined from just two parameters. We impose $\sum_{m = 0}^{M} p_{m} = 1 \quad \quad$ and $\sum_{m = 1}^{M} p_{m} = \phi(T)$ where $\phi(T)$ is the expected proportion of particles to be in icosahedra at that temperature as shown in Fig. \ref{figsAngellAndPhiAll} inset [and Fig. \ref{figP1Phi}(b)]. \textcolor{black}{In the high temperature Arrhenius regime $T \geq 1$, and at slightly lower temperatures ($T \geq 0.7$), the mesocluster size distribution is well described by a decaying exponential $p_m = a^{m-1}p_1$. The parametrisation is discussed in section \ref{sectionPopulationDynamicsParameters}.}

%\textcolor{black}{See Fig. \ref{figProbParams}(a).}

Upon cooling, at around $T\approx0.6$ the number of icosahedra is sufficiently large that the mesoclusters form a percolating network (see Fig. \ref{figFractalDimension}; the change in \textcolor{black}{slope} indicates percolation at mesoclusters with $\gtrsim 500$ particles, which corresponds to $m \gtrsim 70$). Now this percolation does \emph{not} correspond to arrest, because the icosahedra have a limited lifetime \cite{malins2013jcp}. Indeed, the Angell plot in Fig. \ref{figsAngellAndPhiAll} shows no significant feature when the LFS begin to percolate. However percolation leads to a peak in the mesocluster size distribution, which necessitates some explicit considerations for the population dynamics model. We introduce a Gaussian-like weighting function, $W_m(T)$ to account for the peak that forms in the distribution when percolation occurs. $W_m(T)$ is a system-size dependent parameter that controls the location and width of the distribution peak constrained such that the largest mesocluster does not exceed $M$. The steady state solution in the percolated regime is then $p_{m} = a(T)W_m(T)p_{m-1}$.

\textcolor{black}{To describe the dynamics, given the population of icosahedra, we proceed as follows.}
From the mesocluster size distributions, we can determine the super-Arrhenius contribution to $\tau_\alpha$:

\begin{equation}
\tau_\alpha = \tau_\alpha^\mathrm{Arr} \sum_{m}l_m(T)p_m(T)
\label{eqFavourite}
\end{equation}

\noindent \textcolor{black}{Here $\tau^\mathrm{Arr}$ is the relaxation time assuming Arrhenius behaviour, extrapolated from the high-temperature $T>1$ value.}
Each icosahedron is categorised according to the largest mesocluster it joins during its lifetime. \textcolor{black}{The lifetime of an icosahedron is determined by the amount of (simulation) time that has elapsed between the first and last instance of an icosahedron being identified by the TCC.} The mesocluster lifetimes, $l_m$, are the average lifetimes of the icosahedra in the corresponding size category.

The expression above is based on the following assumptions: (1) the dynamics of each particle in the system is represented democratically and (2) particles not in icosahedra have Arrhenius dynamics. We therefore attribute all super-Arrhenius behavior to the emergence of the icosahedra. As noted above this is motivated by the correlation between icosahedra and fragility in model \cite{coslovich2007,royall2014} and metallic  glassformers \cite{shen2009}. Given the populations of the mesoclusters as a function of temperature and extrapolating the dynamical trends we see, we predict $\tau_\alpha$ at temperatures far beyond those accessible to simulation (see Fig. \ref{figsAngellAndPhiAll}).

Since the emergence of the population of icosahedra is associated with super-Arrhenius dynamics,  we consider the dynamical behavior of the system by comparison with an Arrhenius relaxation time $\tau_{\alpha}^\mathrm{Arr}$ which we assume the system would have were no icosahedra to form. Fig. \ref{figFractalDimension}(b) shows the lifetime distribution of all icosahedra whose lifetimes are $\geq0.1\tau_{\alpha}$ across a range of temperatures ($0.58 \leq T \leq 2.5$) scaled by the Arrhenius timescale $\tau_{\alpha}^{\mathrm{Arr}}$. The lifetime distributions collapse onto each other at temperatures $T \gtrsim 1$ but spread out at $T \lesssim 1$.

The mesocluster lifetimes are modeled as a function of size (number of participating icosahedra) for a range of temperatures. We fit the mesocluster lifetimes with two linear expressions; a $T$-dependent expression for small $m$ and a $T$-independent expression for large $m$.

\begin{equation}
l_{m} = 
\begin{cases}
10^{k_{1}(T)m+k_{0}(T)} & \quad \mathrm{for}  \hspace{12pt} m \leq m^* \\
10^{h_{1}m+h_{0}} & \quad \mathrm{for} \hspace{12pt} m > m^*
\end{cases}
\label{eqLifetime1}
\end{equation}

\noindent Here, $m^*$ is the point at which the two expressions are equal, \emph{i.e.} the (feasible) solution to $(k_{1}(T)m+k_{0}(T)) - (h_{1}m+h_{0}) = 0$. ($h_0$ and $h_1$ are listed in table I. \textcolor{black}{and $k_0(T)$ and $k_1(T)$ are described in the following section}) At high temperatures, the mesocluster lifetimes are dominated by the temperature-dependent expression corresponding to $m \leq m^*$. At low temperatures, the mesoclusters are able to grow large enough to pass this ``threshold'' into a regime where the mesocluster lifetimes are no longer related to the temperature and are functions of size only.

\subsection{Mesocluster lifetime parameters}

Throughout the model descriptions,  we utilise a smoothing function with the following form:

\begin{equation}
B_{i}(j) = 0.5 \Big(1 + \mathrm{tanh}[Y(j - X)] \Big)
\label{eqSmooth}
\end{equation}

\noindent where $i$ is a function indicator, $j$ is the variable and $X, Y$ are fitted values. All functions and values are listed in table III.

\begin{figure}%[!htb]
\begin{center}
\includegraphics[width=80mm]{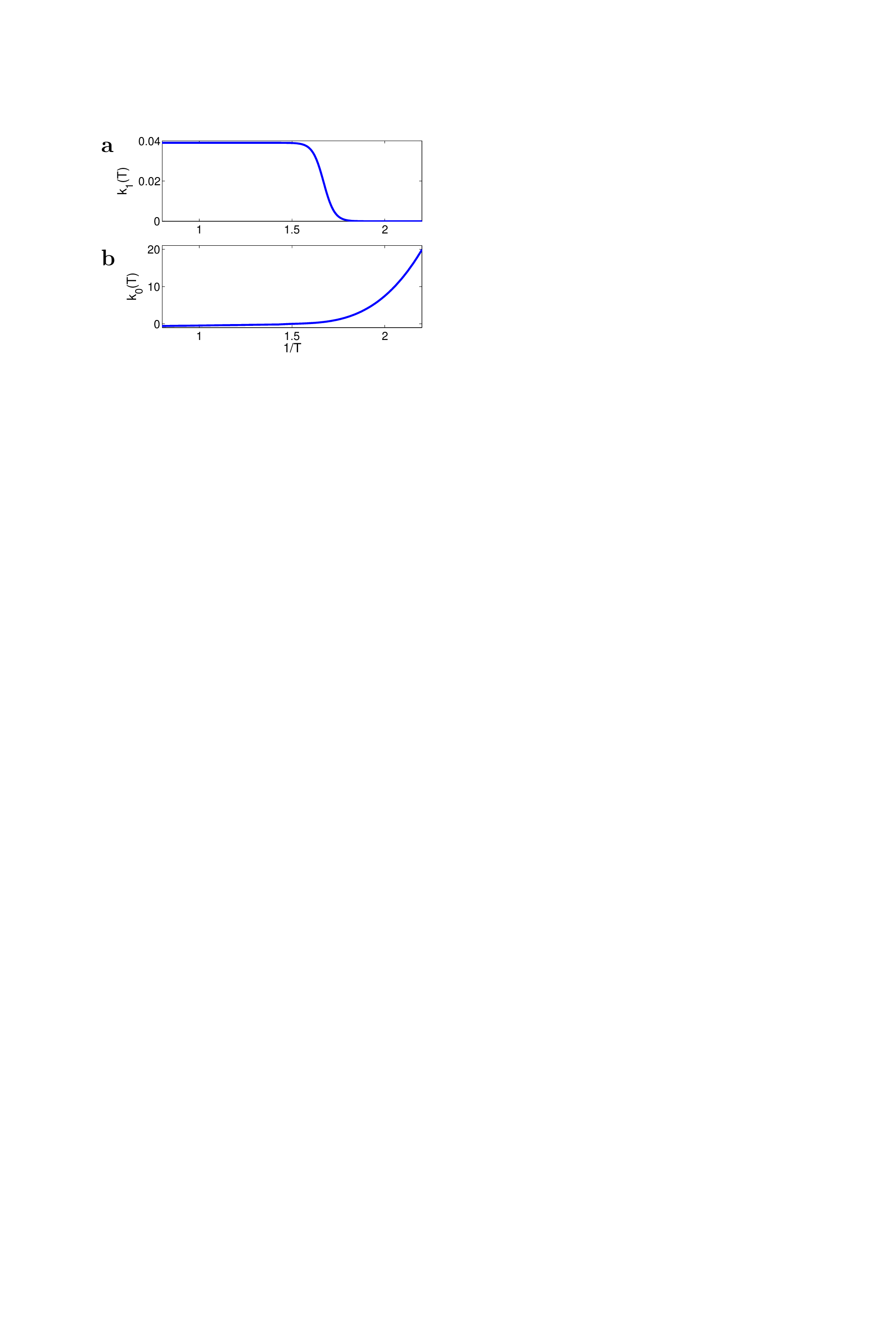}
\caption{The parameters (a) $k_1$ and (b) $k_0$ that describe the \textcolor{black}{slope} and intercept (respectively) of the mesocluster lifetimes in the T-dependent regime.}
\label{figLifeParams}
\end{center}
\end{figure}

The temperature-dependent regime of the lifetime model has two parameters described as follows:

\begin{equation}
k_0(T) = d_3 T^{-3} + d_2 T^{-2} + d_1 T^{-1} + d_0
\label{eqLifetime2}
\end{equation}

\begin{align}
k_1(T) = 
\begin{cases}
0.039 B_{k_1}(\frac{1}{T}) & \quad \mathrm{for} \hspace{12pt} k_0(T) < h_0 \\
0  & \quad \mathrm{for} \hspace{12pt} k_0(T) \geq h_0
\end{cases}
\end{align}

\noindent The coefficients for $d_i$ in the expression for $k_0$ have different values for the two ranges $T>0.7$ and $T \leq 0.7$. These are listed in table II.

\begin{table}[!h]
\begin{center}
\begin{tabular}{| c | c | c |}
\hline
$N$ & $h_0$ & $h_1$ \\
\hline
1372 & 0.0232 & 0.3733 \\
10976 & 0.0033 & 0.8423 \\
87808 & 0.00047 & 1.2 \\
\hline
\end{tabular}
\caption{Fitted parameter values for lifetime model (Eq. \ref{eqLifetime1}).}
\end{center}
\end{table}

\begin{table}[!h]
\begin{center}
\begin{tabular}{| c | c | c | c | c |}
\hline
Region & $d_0$ & $d_1$ & $d_2$ & $d_3$ \\
\hline
$k_0(T), T > 0.7$ & -1.0256 & 0.5033 & 0.0733 & 0 \\
$k_0(T), T \leq 0.7$ & -202.77 & 398.73 & -262.4 & 57.8 \\
\hline
\end{tabular}
\caption{Fitted parameter values for lifetime model (Eq. \ref{eqLifetime2}).}
\end{center}
\end{table}

%\textcolor{black}{\subsection{Mesocluster population dynamics}}
\subsection{Mesocluster population dynamics parameters}
\label{sectionPopulationDynamicsParameters}

\begin{figure}%[!htb] 
\begin{center}
\includegraphics[width=80mm]{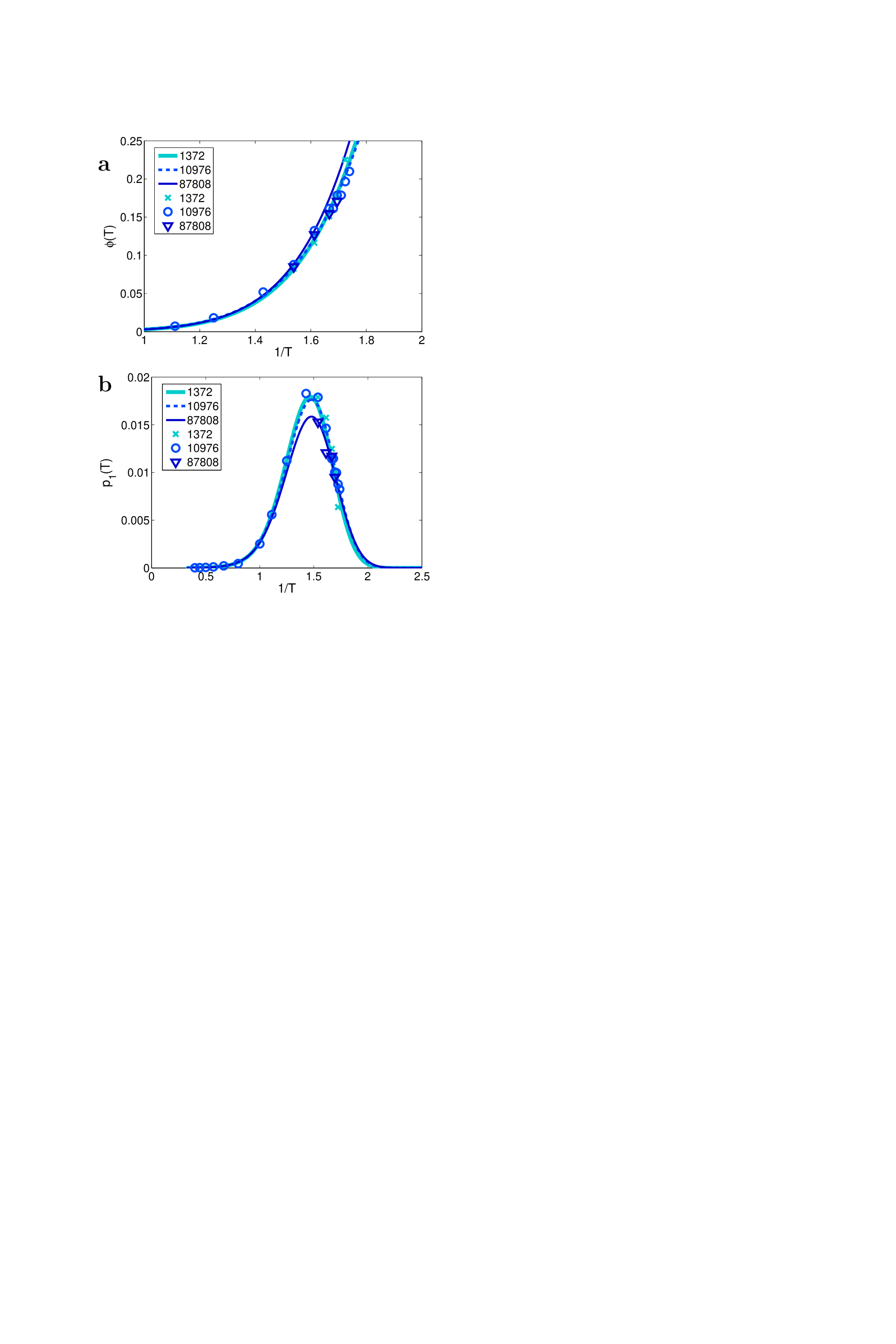}
\caption{(Color online) Data (points) and model descriptions (lines) of (a) the population of particles in icosahedra, $\phi$ and (b) $p_1$. Note that these models are very similar regardless of system size.}
\label{figP1Phi}
\end{center}
\end{figure}

\begin{figure}[!htb]
\begin{center}
\includegraphics[width=80mm]{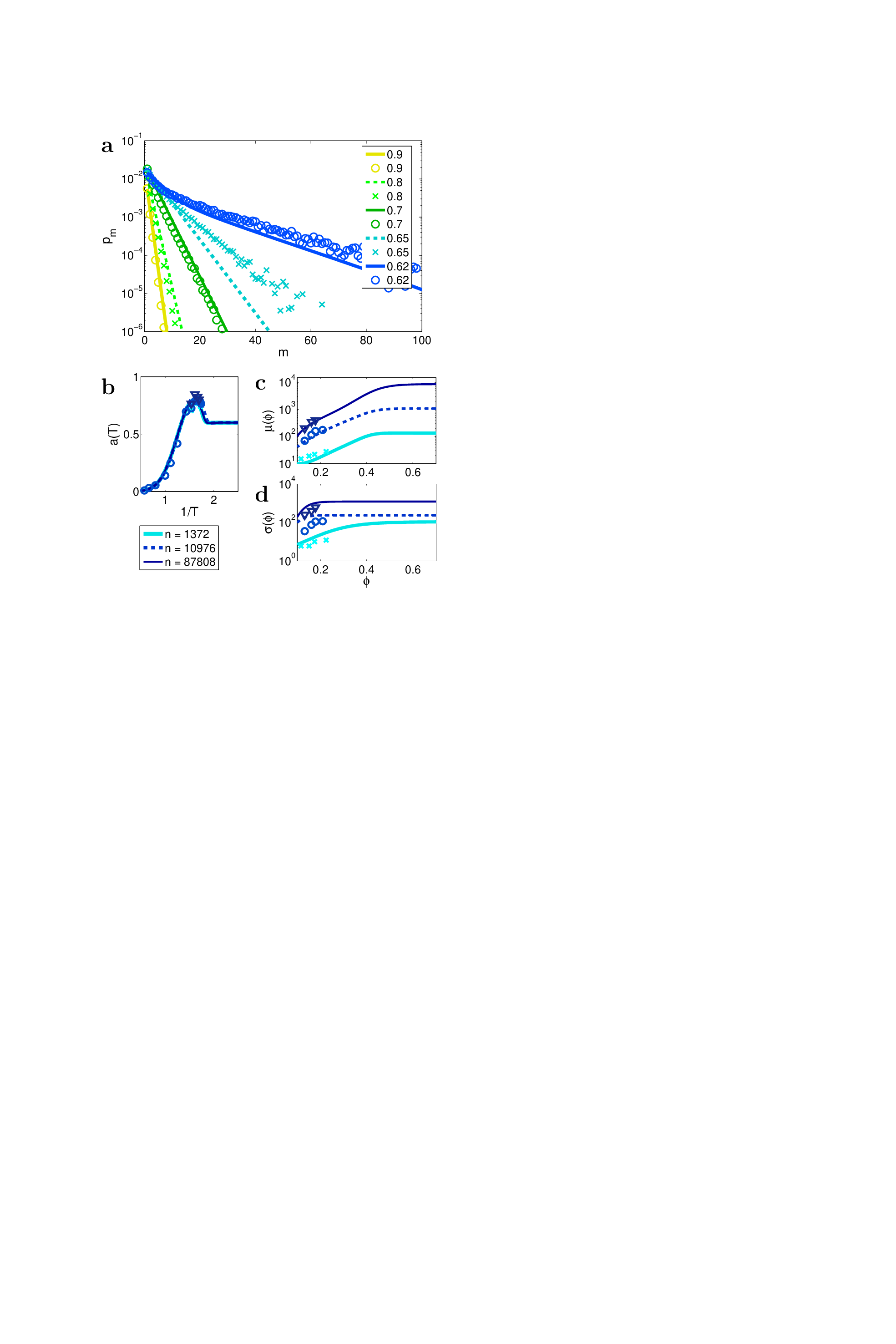}
\caption{(Color online) \textcolor{black}{(a) High temperature data fitted with an exponential decay as described in Eq. \ref{eqSSSolution}.} Model description of (b) the decay parameter, (c) the peak location (``mean'' mesocluster size) parameter, $\mu$ and (d) peak width (``standard deviation'') parameter, $\sigma$ for different system sizes.}
\label{figProbParams}
\end{center}
\end{figure}

Our model for $\phi$ is shown in Fig. \ref{figP1Phi}(a), and described as follows (coefficients $d_i$ listed in table IV):

\begin{align}
\phi(T) &= B_{\phi}\bigg(\frac{1}{T} \bigg) \mathrm{exp} \bigg( d_2 T^{-2} + d_1 T^{-1} + d_0 \bigg) \nonumber \\
&\quad \quad \quad + \phi_\mathrm{max} \Bigg[1 - B_{\phi}\bigg(\frac{1}{T} \bigg) \Bigg] 
\end{align}

\noindent where $B_{\phi}(1/T)$ controls the transition to a plateau. Using simulation data on the number of particles that comprise a typical mesocluster of (arbitrary) size $m$, we infer a linear relation between the proportion of  particles in icosahedra, $\phi$, and the size of mesocluster that could be formed if all the particles in icosahedra were to aggregate into a single mesocluster. We use this linear relation as our model for $M$:

\begin{equation}
M(\phi) = \frac{1}{7.66} \Big(N \phi - 7.1 \Big)
\end{equation}

\noindent where $N$ is the total number of particles in the system (1372, 10976, 87808).

Our model for $p_1$ is shown in Fig. \ref{figP1Phi}(b) and described below (coefficients $d_i$ listed in table IV):

\begin{equation}
p_1(T) = \mathrm{exp} \Big(d_3 T^{-3} + d_2 T^{-2} + d_1 T^{-1} + d_0 \Big)
\end{equation}

\noindent The decay parameter $a$ is plotted in Fig. \ref{figProbParams}(a) and described as follows (coefficients $d_i$ listed in table IV):

\begin{align}
a(T) &= B_a(T) \mathrm{exp} \bigg(d_2 T^{-2} + d_1 T^{-1} + d_0 \bigg) \nonumber \\
&\quad \quad \quad + 0.6\Big( 1 - B_a(T) \Big)
\end{align}

\noindent where $B_a(T)$ is a smoothing function (Eq. \ref{eqSmooth}).

\textcolor{black}{The maximum in $a(T)$ occurs at percolation. Beyond this ($T \lesssim 0.6$), the system is dominated by the large percolating mesocluster (large $m$) and small-mesocluster (small $m$) effects become increasingly negligible. In this simulation inaccessible regime, we assign a fixed value to $a(T)$ for simplicity.}

The function $W_m(T)$ has a Gaussian-like form:

\begin{equation}
W_m(T) = 1 + B_W(m) G \mathrm{exp} \Bigg(- \frac{(m - \mu)^2}{2\sigma^2} \Bigg) 
\end{equation}

\noindent where $\mu$ and $\sigma$ are temperature-dependent parameters that control the location and width of the peak in the distribution, and $G$ is chosen for each $T$ in order to satisfy $\sum_{m = 1}^{M} p_{m} = \phi(T)$. The parameters $\mu$ and $\sigma$ only apply to $T<0.7$, and are only defined in this range.

The parameters $\mu$ and $\sigma$ are given by (coefficients $d_i$ listed in table IV):

\begin{align}
\mu(\phi) &= B_{\mu}(\phi) \Big(d_2 \phi^2 + d_1 \phi + d_0 \Big) \nonumber \\
&\quad \quad \quad + \bigg(\frac{0.75N - 7.1}{7.66} \bigg) \Big(1 - B_{\mu}(\phi) \Big) \\
\sigma(\phi) &= \mathrm{exp} \bigg(d_0 B_{\sigma}(\phi) + d_1\bigg) \\ \nonumber
\end{align}

\noindent with smoothing functions $B_{\mu}(\phi)$ and $B_{\sigma}(\phi)$, and $N$ the total number of particles in the system.

\begin{table}[!h]
\begin{center}
\begin{tabular}{| c | c | c | c |}
\hline
$B_{i}(j)$ & $N$ & $X$ & $Y$ \\
\hline
$B_{\phi}(1/T)$ & 1372 & 1.95 & -30 \\
 & 10976 & 1.95 & -30 \\
 & 87808 & 1.9 & -30 \\
\hline
$B_{a}(T)$ & 1372 & 0.56 & 50 \\
 & 10976 & 0.54 & 50 \\
 & 87808 & 0.55 & 50 \\
\hline
$B_{\mu}(\phi)$ & 1372 & 0.41 & -15 \\
 & 10976 & 0.42 & -12 \\
 & 87808 & 0.43 & -10 \\
\hline
$B_{\sigma}(\phi)$ & 1372 & 0.14 & 5.3 \\
 & 10976 & 0.06 & 20 \\
 & 87808 & 0.07 & 17 \\
\hline
$B_{W}(m)$ & All & 10 & 0.1 \\
\hline
$B_{g_1}(T)$ & All & 1.67 & -17 \\
\hline
\end{tabular}
\caption{Fitted values for smoothing function (Eq. \ref{eqSmooth}) parameters relating to population model components.}
\end{center}
\end{table}

\begin{table}[!h]
\begin{center}
\begin{tabular}{| c | c | c | c | c | c |}
\hline
Fctn. & $N$ & $d_0$ & $d_1$ & $d_2$ & $d_3$ \\
\hline
$\phi(T)$ & 1372 & -14.706 & 10.543 & -1.696 &  \\
 & 10976 & -14.922 & 11.075 & -1.938 &  \\
 & 87808 & -14.697 & 10.482 & -1.627 &  \\
\hline
$p_1(T)$ & 1372 & -12.551 & 1.588 & 9.737 & -4.673 \\
 & 10976 & -12.932 & 3.053 & 8.079 & -4.109 \\
 & 87808 & -13.366 & 4.678 & 6.330 & -3.568 \\
\hline
$a(T)$ & 1372 & -10.718 & 13.104 & -4.091 &  \\
 & 10976 & -10.642 & 12.914 & -3.991 &  \\
 & 87808 & -10.750 & 13.182 & -4.129 &  \\
\hline
$\mu(\phi)$ & 1372 & 18.546 & -165.35 & 794.51 &  \\
 & 10976 & 16.780 & -99.817 & 3439 &  \\
 & 87808 & -173.14 & 2537 & 1013.9 &  \\
\hline
$\sigma(\phi)$ & 1372 & 4.5 & 0.18 &  &  \\
 & 10976 & 5.3 & 0.17 &  & \\
 & 87808 & 7 & 0.1 &  &  \\
\hline
\end{tabular}
\caption{Fitted parameter values for all population model components.}
\end{center}
\end{table}

%\textcolor{black}{\section{Results}} 
\section{Results}
\label{sectionResults}

Our approach is based on the observation that at high temperature there is Arrhenius behavior in the dynamics and very few icosahedra, but at the onset temperature ($T_\mathrm{on}\approx1$ \cite{malins2013jcp}), there is a crossover to super-Arrhenius behavior which is accompanied by the emergence of a population of icosahedra found in mesoclusters which grow upon supercooling \cite{dzugutov2002,coslovich2007,malins2013jcp}. These two dynamical regimes are indicated in the Angell plot in Fig. \ref{figsAngellAndPhiAll} and the population of icosahedra is shown in the inset.

\begin{figure}%[!htb]
\begin{center}
\includegraphics[width=80mm]{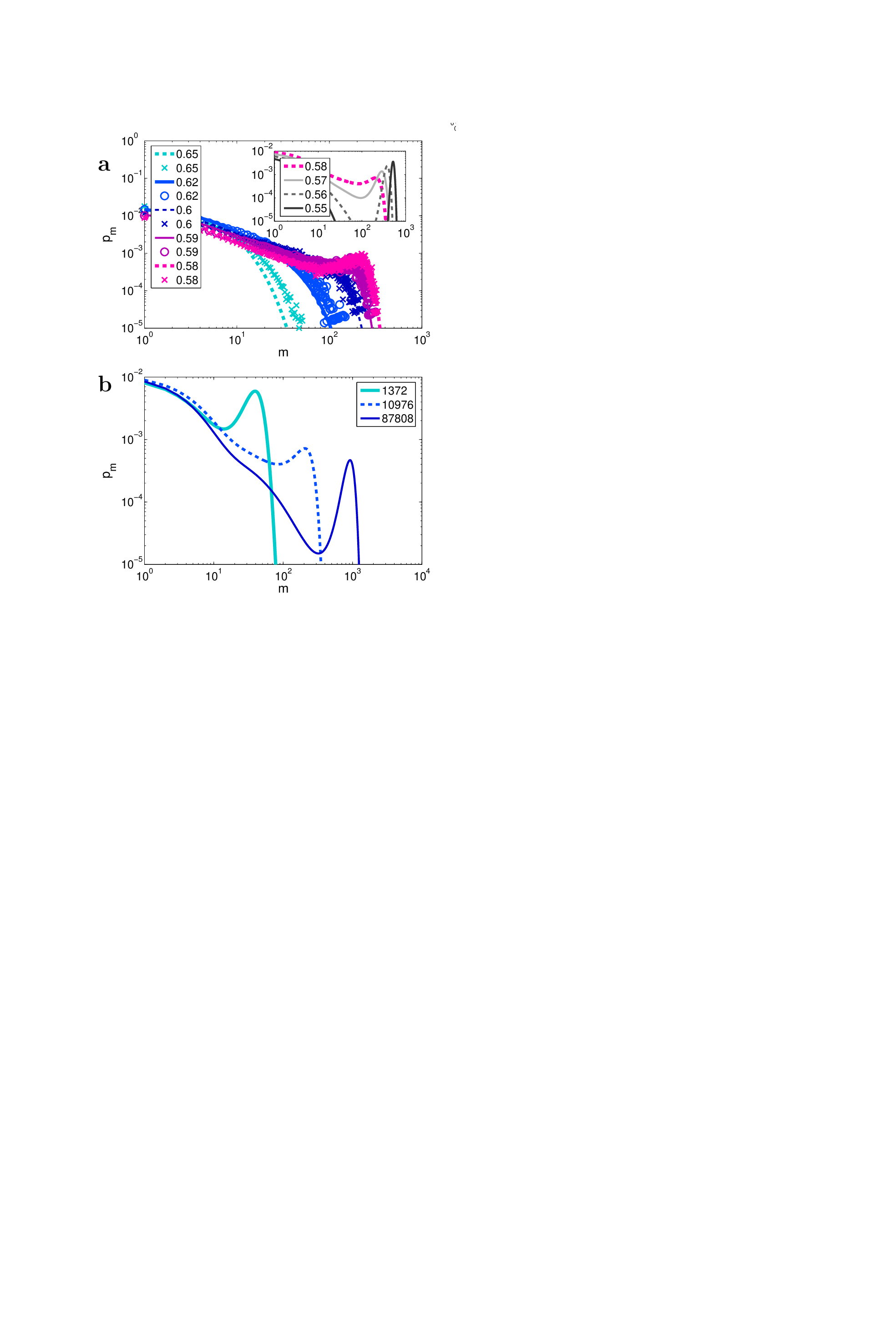} 
\caption{(Color online) (a) Mesocluster probability distributions. Simulation data (points) fitted with the probability model (lines) as descibed in equation \ref{eqMaster}. Inset: the predicted distributions for some temperatures inaccessible to simulations. (b) The probability distribution for different system sizes at a fixed temperature ($T = 0.58$). The decay parameter is almost unchanged across system sizes, but $\mu, \sigma$ and consequently $p_m$ have strong system size dependence. However, note that the distribution shapes are, approximately, compressed/stretched versions of each other.}
\label{figProbs}
\end{center}
\end{figure}

\begin{figure}%[!htb]
\begin{center}
\includegraphics[width=80mm]{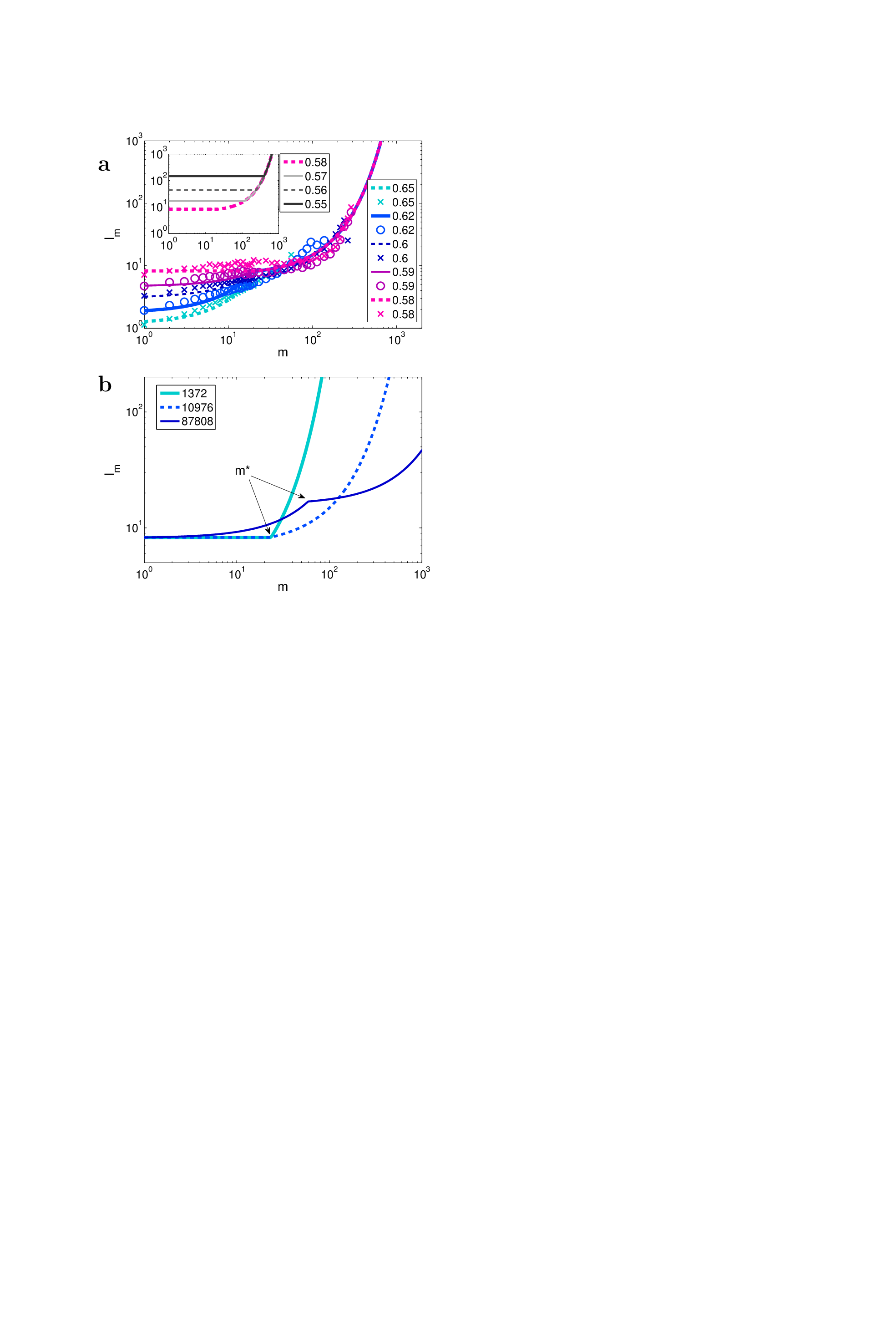}  
\caption{(Color online) (a) Mesocluster lifetimes. Simulation data (points) fitted with the lifetime model (lines) as described in equation \ref{eqLifetime1}. Small mesoclusters (lower values of $m$) have highly $T$-dependent lifetimes, but at some large enough mesocluster size, the lifetimes become $T$-independent and are determined only by the mesocluster size. Inset: the predicted mesocluster lifetimes for some temperatures inaccessible to simulations. (b) The resulting lifetime model for a fixed temperature, $T = 0.58$, for different system sizes. The sudden increase in \textcolor{black}{slope} in (b) occurs at the point $m^*$, where the T-dependent description crosses to the T-independent curve.}
\label{figLife}
\end{center}
\end{figure}

\begin{figure}%[!htb]
\begin{center}
\includegraphics[width=80mm]{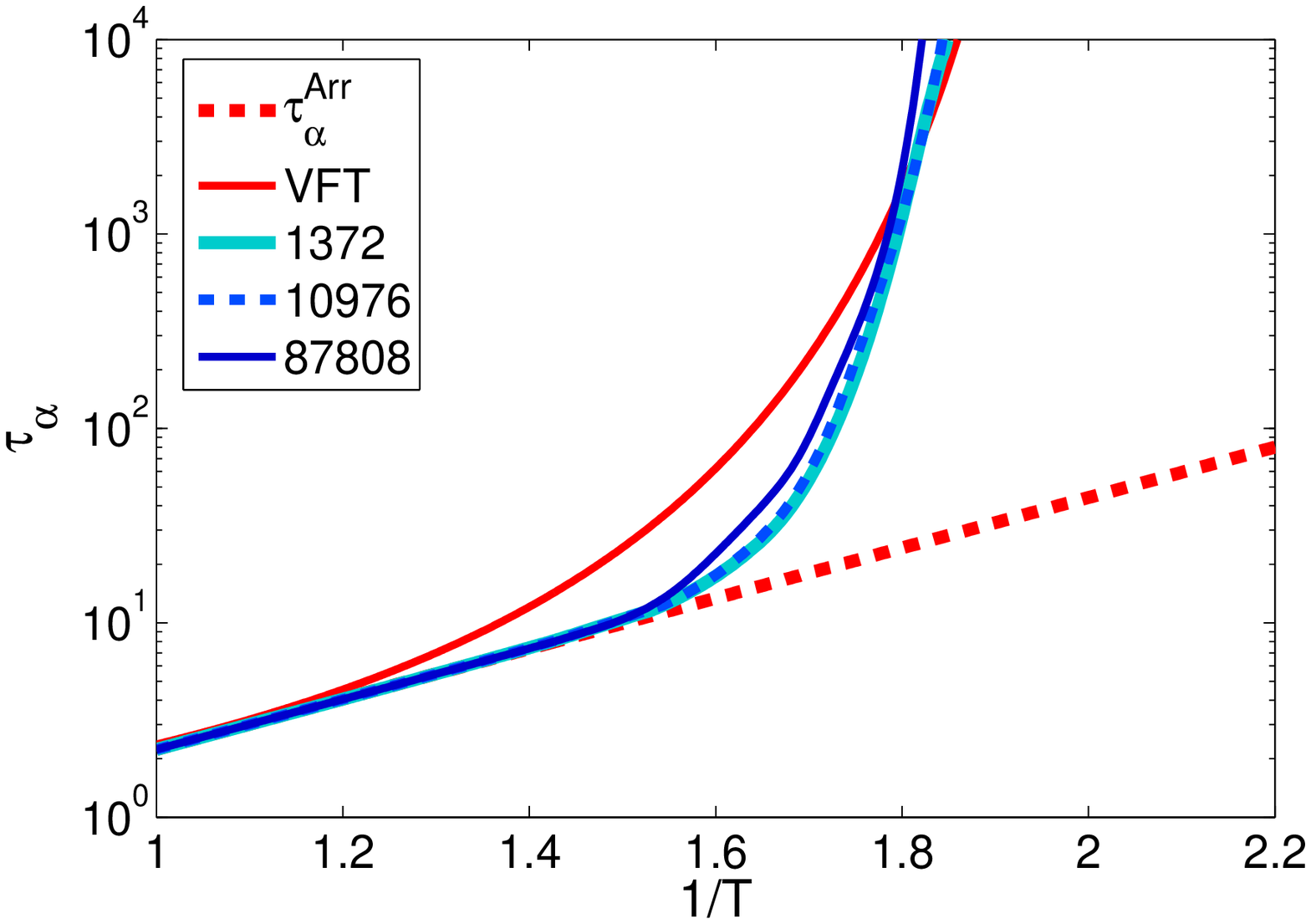}  
\caption{(Color online) The $\alpha$ relaxation time as predicted by the population dynamics model across the three different system sizes. The system size dependence in $p_m$ and $l_m$ produce only very minor differences the model $\tau_{\alpha}$.}
\label{figAngellSizes}
\end{center}
\end{figure}

Motivated by these observations, we propose that the correlation between icosahedra and super-Arrhenius dynamics continues to lower temperature. This gives us a means to predict the relaxation behavior of the system at arbitrary temperature \textcolor{black}{under the assumptions made in constructing the model (section \ref{sectionModel})}. The dynamical behavior we predict from measurements of cluster lifetimes. \textcolor{black}{The global dynamical behavior is shown in the Angell plot in Fig. \ref{figsAngellAndPhiAll} and discussed in section \ref{sectionSummary}. Below we consider some further dynamical features of the model.}

In Fig. \ref{figProbs}(a) we compare the results of the population dynamics model ($p_{m} = a(T)W_m(T)p_{m-1}$) for the size distribution of mesoclusters with simulation data. In the inset, we show the predicted mesocluster distributions for temperatures inaccessible to simulations. The population distribution model results for different system sizes at a fixed temperature ($T = 0.58$) are shown in Fig. \ref{figProbs}(b). Larger systems allow for larger mesoclusters, resulting in the location of the peak having system-size dependence. However, the systems still percolate at the same temperature ($T \approx 0.6$).

In the simulation accessible regime we can directly measure the dynamical properties of the mesoclusters. These we show in Fig. \ref{figLife}(a) which plots the mean lifetime of icosahedra as a function of mesocluster size. This increases strongly with $m$, while retaining some temperature dependence. The mesocluster lifetime model [Eqs. \ref{eqLifetime1}] for different system sizes (for fixed $T = 0.58$) is shown in Fig. \ref{figLife}(b). \textcolor{black}{This shows the behaviour of the model predictions. Note the cusp-like feature shown in this log-log representation which is due to the meeting point $m^*$ between the functions for large and small $m$ in Eq. \ref{eqLifetime1} described in section \ref{sectionModel}.}

The mesocluster lifetimes differ with system size which we explain as follows. Let us assume that in the thermodynamic limit, each mesocluster size has a fixed lifetime. We imagine that in a simulation box, percolating mesoclusters of a given size correspond to a \emph{distribution} of larger mesoclusters in the thermodynamic limit, rather than being a fixed size. So in the thermodynamic limit these larger mesoclusters would have a distribution of lifetimes, but here we assign a single lifetime for each sampled size. Crucially, this effect varies with system size.

In Fig. \ref{figAngellSizes} we show the structural relaxation time our model predicts for different system sizes across the whole range of $T$ (Eq. \ref{eqFavourite}). Despite the strong system size dependence in both the mesocluster population distributions and lifetimes, the resulting relaxation times are scarcely \textcolor{black}{effected by the} system size.

%\textcolor{black}{\section{Summary and Discussion}} 
\section{Summary and Discussion}
\label{sectionSummary}

We have presented an approach to predict the dynamics of a glass forming liquid at arbitrary temperatures. This we have done by decomposing a model glassformer into an effective system of \textcolor{black}{\emph{mesoclusters of}} locally favored structures \textcolor{black}{whose population is described by}
%. We introduce 
a population dynamics model which we parameterize with results from simulations.
% to predict the population, connectivity and 
\textcolor{black}{The} lifetime of mesoclusters of LFS \textcolor{black}{are also parameterized with simulations} 
%for different system sizes. 
\textcolor{black}{Under the assumptions above and that the super-Arrhenius dynamics can be attributed to the population of particles in mesoclusters of icosahedra, our model predicts dynamical behaviour at arbitrary temperature.} \textcolor{black}{In its present form} the model predicts that there is no thermodynamic \textcolor{black}{glass} transition. \textcolor{black}{We have considered different system sizes and find that } 
while the population and lifetime components of the model are strongly system size dependent, the resulting relaxation time is scarcely dependent on system size.

Whether or not there is a thermodynamic transition, in the sense of a divergence in relaxation timescales at finite temperature, boils down to whether the lifetime of icosahedra in mesoclusters $l_m$, diverges. In Fig. \ref{figLife}(a) we see that it does not. Although our data are compatible with dynamical divergence of $l_m$, --- that is to say $l_m$ can be fitted in the regime accessible to simulation such that it diverges at finite temperature --- better fits are obtained with non-divergent behaviour.

Now since the population dynamics model itself does not exhibit a phase transition, perhaps one could argue that it is natural that we do not find a divergence in relaxation time. One might imagine that population dynamics models in which a phase transition is encountered would lead to dynamical divergence \cite{krapivsky}. It is also possible that further refinement may fit the relaxation time data better than our current approach [Fig. \ref{figsAngellAndPhiAll}], which might \textcolor{black}{provide further insight into} the question as to whether there is a thermodynamic glass transition. At this stage we observe that our model which predicts {\em no} thermodynamic transition actually {\em over-estimates} the super-Arrhenius nature of the dynamics. This would lend support to the observation that within this framework there should be no transition as also found in other treatments of LFS \cite{tarjus2005,eckmann2008,lerner2009}.

We also plot in Fig. \ref{figsAngellAndPhiAll} predications from geometric frustration \cite{tarjus2005}. Here $\tau_\alpha(T)=\tau_\infty \exp \left( \Delta E^*(T)+E_\infty/k_BT \right)$ where $E_\infty$ is the Arrhenius contribution. \textcolor{black}{Below} the onset temperature $T_\mathrm{on}$ the super-Arrhenius contribution $ \Delta E=0$, for $T<T_\mathrm{on}$, $\Delta E(T)=B k_BT_c \left(  1-\frac{T}{T_\mathrm{on}}  \right)^\psi$ where $B=650$, $\psi=8/3$ and $T_c=0.65$. We see that the results, also predicated on icosahedra, seem to describe the dynamical behavior in a similar way to our model. It is possible that certain aspects of geometric frustration are captured by our approach.

One explanation for the rather strong increase in $\tau_\alpha$ exhibited in Fig. \ref{figsAngellAndPhiAll} might be that our model doesn't include mesocluster scission \textcolor{black}{or coalescence}, because the mesocluster size increases/decreases only by one. Larger changes in mesocluster size might lead to closer agreement with simulation data, but would not change the picture in a qualitative fashion. System sizes for simulations which represent certain properties of deeply supercooled systems are small \cite{berthier2013overlap,hedges2009,speck2012}, but it is tantalizing to consider parameterizing the model with such data. Alternatively one can consider how the network geometry might be influenced by certain scaling properties near an assumed transition \cite{stevenson2006}.

Our model underestimates the initial %super-Arrhenius 
increase in structural relaxation time in the dynamical regime where it is well described by MCT (Fig. \ref{figsAngellAndPhiAll}). It is tempting to suggest that this is related to our emphasis on icosahedra in describing the dynamical slowdown. In the temperature regime in question, \textcolor{black}{LFS (icosahedra)} are relatively few in number and there is no percolating network. Moreover it is possible that in this regime, dynamical slowdown may be dominated by lower-order correlations such as the 2-body, 3-body etc. \textcolor{black}{These arguments are supported by Banerjee \emph{et al.} \cite{banerjee2014} and Nandi \emph{et al.} \cite{nandi2015} who have suggested that two-point based relaxation times strongly increase before higher-order contributions. In Fig. \ref{figsAngellAndPhiAll} we show the MCT fit which describes the simulation data accurately for $0.7\gtrsim T\gtrsim0.6$ but diverges at $T_\mathrm{MCT}=0.57$ \cite{lacevic2003}.} One imagines that better agreement might be obtained by including contributions from MCT in this dynamical range, noting that these can be systematically extended at least to 4th order \cite{janssen2015}. At deeper supercoolings where MCT diverges, our population dynamics model presumably captures other dominant relaxation pathways absent from MCT \cite{cavagna2012}, at least for the Wahnstr\"{o}m model. Thus it is in the deeply supercooled regime beyond MCT whose dynamics remain inaccessible to the particle-resolved techniques of computer simulation and colloid experiment where our approach has most to contribute. \textcolor{black}{Other possibilities to explain the discrepancy at weak supercooling ($T>T_\mathrm{MCT}$) include amorphous order distinct from icosahedra. The correlation of the dynamics with the icosahedra is high in the Wahnstr\"{o}m model, but it is not perfect \cite{hocky2014}. Considerations from other work \cite{widmercooper2006,jack2014} suggest that other contributions from the structure may also contribute to the dynamics \cite{royall2015physrep}.}

Before closing, we comment on the generality of our results. Recently, \cite{hocky2014} a number of models have been compared. Of those, the Wahnstr\"{o}m model considered here exhibits the strongest correlations between structure and dynamics, so one might expect this to be most likely to exhibit a structure-based transition at finite temperature. That our analysis hints towards no such transition therefore suggests the same should hold at least for the range of models considered \cite{hocky2014}. Our approach may also be used to optimise metallic glassformers such as Cu$_x$Zr$_{1-x}$. Like the Wahnstr\"{o}m model, these materials are well known to exhibit correlations between non-Arrhenius dynamics and the emergence of icosahedra \cite{cheng2011}. \textcolor{black}{These models also exhibit networks of icosahedra, whose emergence seems similarly related to the crossover to super-Arrhenius dynamics as the Wahnstr\"{o}m model we consider \cite{soklaski2013,soklaski2015}.}

\subsection*{Acknowledgements} 
The authors would like to thank Andrea Cavagna, Daniele Coslovich, Jens Eggers, Bob Evans, Rob Jack, Gilles Tarjus and Francesco Turci for many helpful discussions.
CPR would like to acknowledge the Royal Society for financial support and the European Research Council under the FP7 / ERC Grant agreement n$^\circ$ 617266 ``NANOPRS''.
RP is funded by EPSRC grant code EP/E501214/1. This work was carried out using the computational
facilities of the Advanced Computing Research Centre, University of Bristol.

\bibliography{icos}

\end{document}